\documentclass[prb,twocolumn,showpacs,aps,superscriptaddress,floatfix]{revtex4-1}
\usepackage{amsmath}
\usepackage{amssymb}
\usepackage{bm}
\usepackage{graphicx}
\usepackage{color}
\usepackage{hyperref}
\usepackage{verbatim}
\graphicspath{{Figs/}}
\begin{document}
    \title{Bulk spin conductivity of three-dimensional topological insulators}
    \author{R.S. Akzyanov}
        \affiliation{Dukhov Research Institute of Automatics, Moscow, 127055 Russia}

    \begin{abstract}
    We study the spin conductivity of the bulk states of three-dimensional topological insulators within Kubo formalism. Spin Hall effect is the generation of the spin current that is perpendicular to the applied voltage. In the case of a three-dimensional topological insulator, applied voltage along $x$ direction generates transverse spin currents along $y$ and $z$ directions with comparable values. We found that finite non-universal value of the spin conductivity exists in the gapped region  due to the inversion of bands. Contribution to the spin conductivity from the vertex corrections enhances the spin conductivity from the filled states. These findings explain large spin conductivity that has been observed in topological insulators.
    \end{abstract}
    
    
    \maketitle
\section{Introduction} 
Generation of the large spin currents is of interest for the fast and non-volatile magnetic memory\cite{Ralph2008,Kent2015}. Such spin current can be generated in systems with time-reversal symmetry without the spin Hall effect\cite{Dyakonov1971,Hirsch1999,Sinova2015}. 
Various materials with the strong spin-orbit coupling have been proposed as a platform for a spin Hall effect\cite{Murakami2003,Sinova2004,Kane2005}. However, it occurs that for materials with the Rashba coupling spin current is small due to vertex corrections~\cite{Inoue2004,Raimondi2005,Dimitrova2005}.
Good material for the spintronic applications should have high spin conductivity with low charge conductivity. Since spin conductivity obeys time-reversal symmetry finite spin conductivity can be realized even for the insulating state where charge conductivity is suppressed\cite{Murakami2004}. Such materials - spin Hall insulators - are an ideal charge to spin converters. 

Topological insulators have a topologically non-trivial band structure that results in a robust surface states\cite{Hasan2010}. 
Recently, record spin currents have been measured in topological insulators such as Bi$_2$Se$_3$~\cite{Mellnik2014} and Bi$_{0.1}$Sb$_{0.9}$~\cite{Khang2018}. The origin of such large spin currents in topological insulators is under debate. Theoretical calculations predict much less value of the bulk spin conductivity for a latter material than it was measured in the experiment\cite{Sahin2015}. It has been proposed that surface states of the topological insulator are responsible for large spin conductivity\cite{Shiomi2014,Wang2015,Wang2016}. Other papers identify bulk states as a major source of spin current in topological insulators~\cite{Jamali2015,Han2017}.
In topological insulators along with the anti damping torque (that is usually associated with the spin Hall currents) damping torque with the comparable amplitude is present~\cite{Mellnik2014}. Usually, damping torque corresponds to the generation of non-equilibrium spin density or Rashba-Edelstein effect~\cite{Edelstein1990,Han2018}. However, three-dimensional topological insulators have inversion symmetry for the bulk states and the Rashba-Edelstein effect should be absent. It means that the presence of the significant damping torque in topological insulators is a puzzle that requires further investigation.

The spin conductivity of the surface states of topological insulators has been studied previously~\cite{Peng2016,Ghosh2018,Akzyanov2019a,Akzyanov2019}. Bulk spin conductivity in topological insulators has been calculated using a simplified tight-binding model without taking into account vertex corrections in Ref.~\onlinecite{Ghosh2018}. 
Previously, we studied the spin conductivity of the surface states of topological insulators. We found that spin conductivity contribution from the surface states is significantly smaller than the spin conductivity measured in the experiment\cite{Akzyanov2019}. We used a simplified one orbital two-dimensional model for the bulk states of the topological insulator that neglects the actual multi-orbital structure of the Hamiltonian of topological insulators and missed a term in a contribution to the spin conductivity from the filled states. This missed term leads to the suppression of the spin conductivity for the bulk states for $\sigma_{xy}^z$ component as it should be for a Rashba Hamiltonian. Multi-orbital structure makes a big difference due to the inversion of bands in topological insulators~\cite{Liu2010}. This inversion leads to the non-trivial topology of the band structure and can affect spin conductivity in the gapped region and vertex corrections as well. Also, additional components of the spin conductivity tensor are allowed in three dimensions in comparison with the two-dimensional models. 

In present work, we study the spin conductivity of the bulk states of three-dimensional topological insulators. We use low energy Hamiltonian of Bi$_2$Se$_3$ that was obtained in Ref.~\onlinecite{Liu2010}. Spin conductivity is calculated within Bastin-Kubo formalism by taking into account random point disorder and vertex corrections. We get that applied voltage generates transverse spin Hall currents in perpendicular directions with comparable amplitudes. We found that finite non-universal value of the spin conductivity is present even if the spectrum is gapped. Spin conductivity contribution from the vertex corrections is comparable to the intrinsic contribution from the filled states and have the same sign. The maximal spin conductivity around $e/(2\pi)^3 \cdot $\AA$^{-1}$ is realized if chemical potential is in the conduction zone near the bandgap. Spin conductivity is suppressed for large doping and is determined by the interplay of the quadratic corrections and leading linear terms. We compare our results with the experimental data.

The paper is organized as follows. In Sec.~\ref{model} we introduce low energy Hamiltonian of the bulk states of topological insulator and model for the calculations of the spin conductivity. In Sec.~\ref{disorder} impurity averaged Green's functions and vertex corrections are calculated. These results are used in Sec.~\ref{fermi_surface} for the calculation of the contribution to the spin conductivity from the states at the Fermi energy. In Sec.~\ref{total} contribution to the spin conductivity from the filled states and total spin conductivity are obtained. In Sec.~\ref{discussion} we compare our results with the experimental data and discuss obtained results.

\section{Model}\label{model}
\begin{figure}[t!]
    \centering
    \includegraphics [width=8.5cm, height=4.5cm]{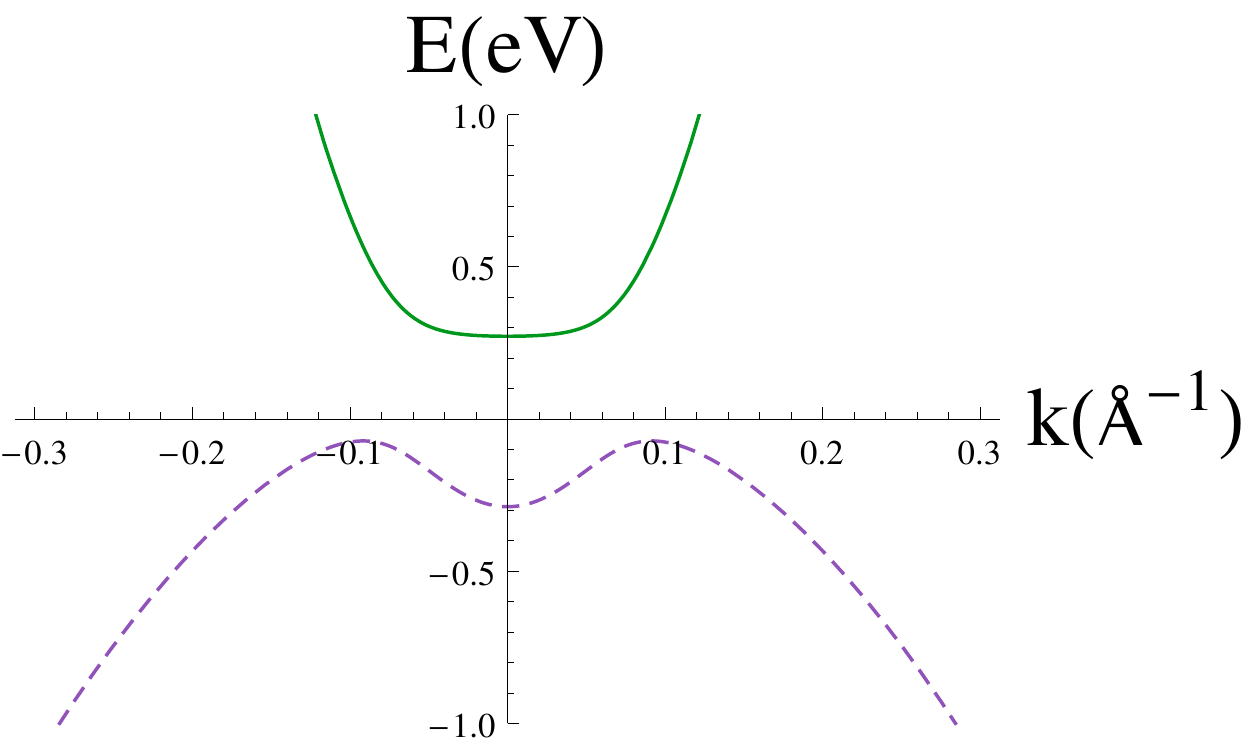}
    \caption{Energy spectrum~\eqref{spectrum} for Bi$_2$Se$_3$ in $k_z,k_y=0$ plane as a function of in-plane momentum $k_x$. Green line corresponds to the conduction band $E_+$, purple dashed line to the valence band $E_-$}
    \label{spectra}
\end{figure}

Low energy Hamiltonian of the bulk states of the three dimensional topological insulators, such as Bi$_2$Se$_3$, is given by~\cite{Liu2010} ($\hbar=1$)
\begin{eqnarray}\label{H0}
\hat{H}\!=C_0+C_1k_z^2+C_2(k_x^2+k_y^2)+\\\nonumber[M_0+M_1k_z^2+M_2(k_x^2+k_y^2)]\tau_z+ \\ \nonumber B_0k_z\tau_y+A_0(k_y\sigma_x-k_x\sigma_y)\tau_x,
\end{eqnarray} 
where $\mathbf{\sigma}=(\sigma_x,\sigma_y,\sigma_z)$ are the Pauli matrices acting in the spin space, $\mathbf{\tau}=(\tau_x,\tau_y,\tau_z)$ are the Pauli matrices acting in the orbital space, $k_x=k\cos \phi$ and $k_y=k\sin \phi$ are the in-plane momentum components, $k_z$ is the out-of-plane momentum. We take values of parameters from Ref.~\onlinecite{Liu2010}: $A_0=3.33$ eV$\cdot $\AA, $B_0=2.26$ eV$\cdot $\AA, $C_0=-0.0083$ eV, $C_1=5.74$ eV$\cdot $\AA$^2$, $C_2=30.4$ eV$\cdot $\AA$^2$, $M_0=-0.28$ eV, $M_1=6.86$ eV$\cdot $\AA$^2$, $M_2=44.5$ eV$\cdot $\AA$^2$. We neglect hexagonal warping. We address effects of warping in Discussion. The spectrum of the Hamiltonian given by Eq.~\eqref{H0} is doubly degenerate
\begin{eqnarray}\label{spectrum}
E_{\pm}=C_0+C_1k_z^2+C_2k^2\pm \\ \nonumber \sqrt{A_0^2k^2+B_0^2k_z^2+M_0^2+M_1^2k_z^4+M_2^2k^4}.
\end{eqnarray}
The spectrum is shown in Fig.~\ref{spectra}. As we can see, there is a gap in the spectrum. Also, this spectrum has significant asymmetry between the valence and conduction bands due to quadratic corrections $C_1$ and $C_2$ to the spectra.

In general, the spin conductivity can be presented as a sum of three terms~\cite{Yang2006,Kodderitzsch2015}
\begin{eqnarray}
\sigma_{{\alpha}\beta}^{\gamma}=\sigma_{{\alpha}\beta}^{ I\gamma}+\sigma_{{\alpha}\beta}^{ II\gamma}+\sigma_{{\alpha}\beta}^{III\gamma},
\end{eqnarray}
where the first two items correspond to a contributions from the states at the Fermi level and the third one from the filled states. Here $\alpha=x,y,z$ corresponds to the direction of flow of spin current, $\beta=x,y,z$ corresponds to the direction of applied voltage, and $\gamma=x,y,z$ denotes the spin projection of the spin current.

At zero temperature $\sigma_{{\alpha}\beta}^{I\gamma}$ and $\sigma_{{\alpha}\beta}^{II\gamma}$ contributions of the states at the Fermi energy can be written as~\cite{Inoue2004,Yang2006}
    \begin{eqnarray}\label{spin_surface}
\sigma_{{\alpha}\beta}^{I\gamma }=\frac {e}{4\pi}\int \frac{d^3{\bf k}}{(2\pi)^3} \textrm{Tr} [j_{\alpha}^{\gamma}\, G^+
\,V_{\beta}  \, G^-],\,\,\\
\label{spin_surface1}
\sigma_{{\alpha}\beta}^{II\gamma }\!=\!-\frac {e}{8\pi}\!\int\! \frac{d^3{\bf k}}{(2\pi)^3}\textrm{Tr} [j_{\alpha}^{\gamma}\, G^+ \,V_{\beta}  \, G^+\! +\! j_{\alpha}^{\gamma}\, G^- \,V_{\beta}  \, G^- ].\,\,\,
\end{eqnarray}
Here $d^3{\bf k}=dk_x\,dk_y\,dk_z$, $j_{\alpha}^{\gamma}= \{\sigma_{\gamma}, v_{\alpha}\}/4$ is the definition of the spin current,  $v_{\alpha}= \partial H /\partial k_{\alpha}$ is the velocity operator, $V_{\alpha}$ is the velocity operator with vertex corrections, $\{\,,\}$ means the anti commutator, and $G^{\pm}$ are the retarded and advanced disorder averaged Green's functions, which will be specified later. Note, that the definition of the spin velocity operator $j_{\alpha}^{\gamma}$ that we use is common and should be used for the calculation of the spin conductivity\cite{Sun2008}.

The contribution to the spin conductivity from the filled states~\cite{Yang2006,Kodderitzsch2015} 
\begin{eqnarray}\label{spin_filled}
\sigma_{{\alpha}\beta}^{III\gamma}=\frac {e}{8\pi}\int \frac{d^3{\bf k}}{(2\pi)^3} \int\limits_{-\infty}^{+\infty} f(E) dE \times\\ \nonumber \textrm{Tr} [j_{\alpha}^{\gamma}\, G^+\,v_{\beta}  \, \frac {d G^+}{dE}-j_{\alpha}^{\gamma}\, \frac {d G^+}{dE}\,v_{\beta}  \,G^+ + c.c.] ,
\end{eqnarray}
 where $f(E)= (1 + \exp [E-\mu]/T)^{-1}$ is the Fermi distribution function (that is Heaviside step for zero temperature), $c.c$ means complex conjugation.  Here Green's function is  $G^{\pm}(E)=[1+\Sigma^{\pm}(E) G_0^{\pm}(E)]^{-1} G_0^{\pm}(E)$, where Green's function in a clean limit is $G_0^{\pm}(E)=(E\pm i0 - \hat{H})^{-1}$ and $\Sigma^{\pm}(E)$ is the self-energy which will be specified latter.
    
\section{Disorder and vertex corrections}\label{disorder}
We will describe disorder by a potential $V_{\textrm{imp}}=u_0 \sum_i \delta(\mathbf{r}-\mathbf{R}_j)$, where $\delta(\mathbf{r})$ is the Dirac delta function, $\mathbf{R}_j$ are the positions of the randomly distributed point-like impurities with the local potential $u_0$ and concentration $n_i$. We assume that the disorder is Gaussian, that is, $\langle V_{\textrm{imp}} \rangle=0$ and $\langle V_{\textrm{imp}}(\mathbf{r}_1) V_{imp}(\mathbf{r}_2) \rangle=n_i u_0^2 \delta (\mathbf{r}_1-\mathbf{r}_2)$. 

In the self-consistent Born approximation (SCBA), the impurity-averaged Green's functions can be calculated as
$
G^{\pm}=G_0^{\pm}+G_0^{\pm}\Sigma^{\pm} G^{\pm},
$
where $G_0^{\pm}=(\mu\pm i0-\hat{H})^{-1}$ are bare retarded/advanced Green's functions of the Hamiltonian~\eqref{H0}
and $\Sigma^{\pm}$ is the self-energy, $\mu$ is the chemical potential. Self-energy is defined as
$\Sigma^{\pm}=  \langle V_{\textrm{imp}} G^{\pm} V_{\textrm{imp}}\rangle.$
In the case under consideration, we can calculate the self-energy $\Sigma^{\pm}=\Sigma'\mp i\Gamma$ using an Dyson equation $\Sigma^{\pm}=n_iu_0^2\sum_k G^{\pm} $.    
The self-energy has a nontrivial structure in the orbital space $\tau$. Along with the trivial in orbital space part of self-energy $\Sigma_0 \tau_0$ it has non-trivial one $\Sigma_z \tau_z$. Therefore, the expression for $G^\pm$ is similar to $G^\pm_0$, in which $\pm i0$ is replaced by $\pm i\Gamma_0$, $\mu$ by $\mu-\Sigma'_0$ and $M_0$ by $M_0-\Sigma'_z\mp i\Gamma_z$. The value $\Gamma_0$ describes the scattering rate while $\Gamma_z$ describes the orbital scattering rate. Thus, the impurity averaged Green's function can be calculated as $G^{\pm}=(1+\Sigma^{\pm} G_0^{\pm})^{-1} G_0^{\pm}$. 

Here we consider that chemical potential lies outside the gap. In this case we can neglect a small correction to the value of $\mu$ due to real part of the self-energy and put $\Sigma' =0$. In this limit we suppose that scattering rates $\Gamma_0, \Gamma_z \rightarrow 0$ are small and we obtain that
\begin{eqnarray}
\Gamma_0=n_i u_0^2\sum_k\,\textrm{Im}\, \textrm{Tr}[G_0^+], \\
\Gamma_z=n_i u_0^2\sum_k\,\textrm{Im} \,\textrm{Tr}[\tau_z G_0^+].
\end{eqnarray}

\begin{figure}[t!]
    \centering
    \includegraphics [width=8.5cm, height=4.5cm]{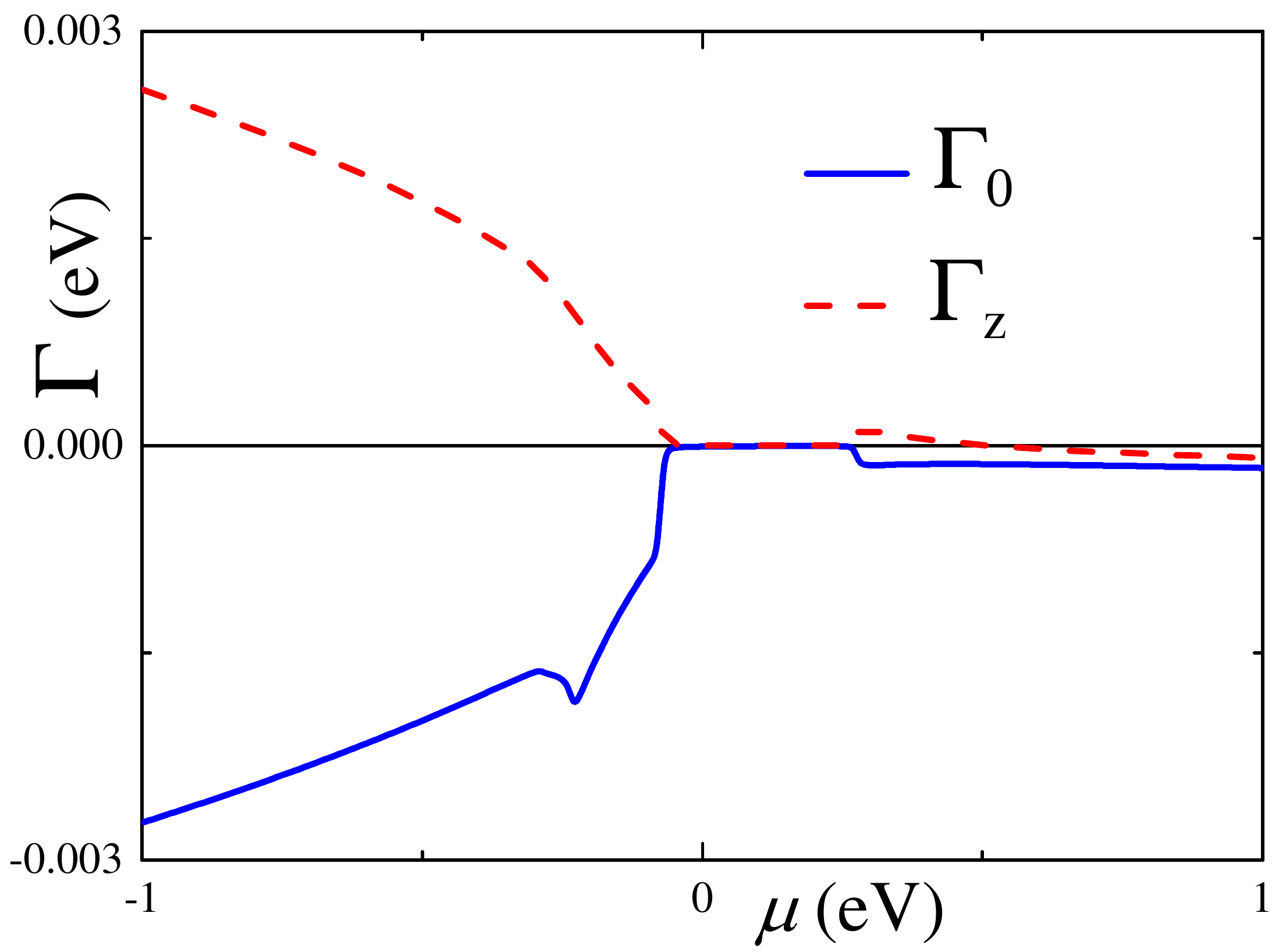}
    \caption{Scattering rates $\Gamma_0$ (blue line) and $\Gamma_z$ (red dashed line) as a functions of a chemical potential $\mu$.}
    \label{scattering}
\end{figure}

Scattering rates are given in Fig.~\ref{scattering}. As we can see impurity scattering is more significant for the valence bands $\mu<0$ than for the conduction bands $\mu>0$.
Orbital scattering is always smaller $|\Gamma_0|>|\Gamma_z|$. Scanning tunneling microscopy study shows that there is approximately 1 defect per $10^{6}$\AA$^3$ for clean samples~\cite{Cheng2010}.
Typical impurity potential we estimate as\cite{DasSarma2015} $u_0=4\pi e^2r_s^2/\kappa \sim 10^2$eV\AA$^3$.
Here screening is taken as $r_s=(a^2c)^{1/3}=7.7$\AA, where $a=4.14 $\AA, $c=28.7$\AA, are the lattice constants, dielectric constant is\cite{Kim2012} $\kappa=100$. Thus, we can estimate $n_i u_0^2 \sim 10^{-2}$eV$^2\cdot$ \AA$^{3}$ that we will use through the text.

In the self-consistent Born approximation, following the approach described in Ref.~\cite{Shon1998}, we can derive an equation for the vertex corrected velocity operator
\begin{eqnarray}\label{vcor}
V_{\alpha}(\mathbf{k})=v_{\alpha}(\mathbf{k}) + n_i u_0^2 \sum\limits_k G^+(\mathbf{k}) V_{\alpha}(\mathbf{k}) G^-(\mathbf{k}).
\end{eqnarray} 

Point-like disorder renormalize $k$-independent part of the velocity operator. Thus, we write down
\begin{eqnarray}\label{hrz}  
V_x=v_x + (A_{0c}-A_0)\sigma_y\tau_x, \\ \nonumber
V_y=v_y - (A_{0c}-A_0)\sigma_x\tau_x,\\ \nonumber
V_z=v_z - (B_{0c}-B_0)\tau_y,
\end{eqnarray}
where velocity operators with vertex corrections are calculated by the substitution of Eq.~\eqref{hrz} into Eq.~\eqref{vcor}. 

\begin{figure}[t!]
    \centering
    \includegraphics [width=8.5cm, height=4.5cm]{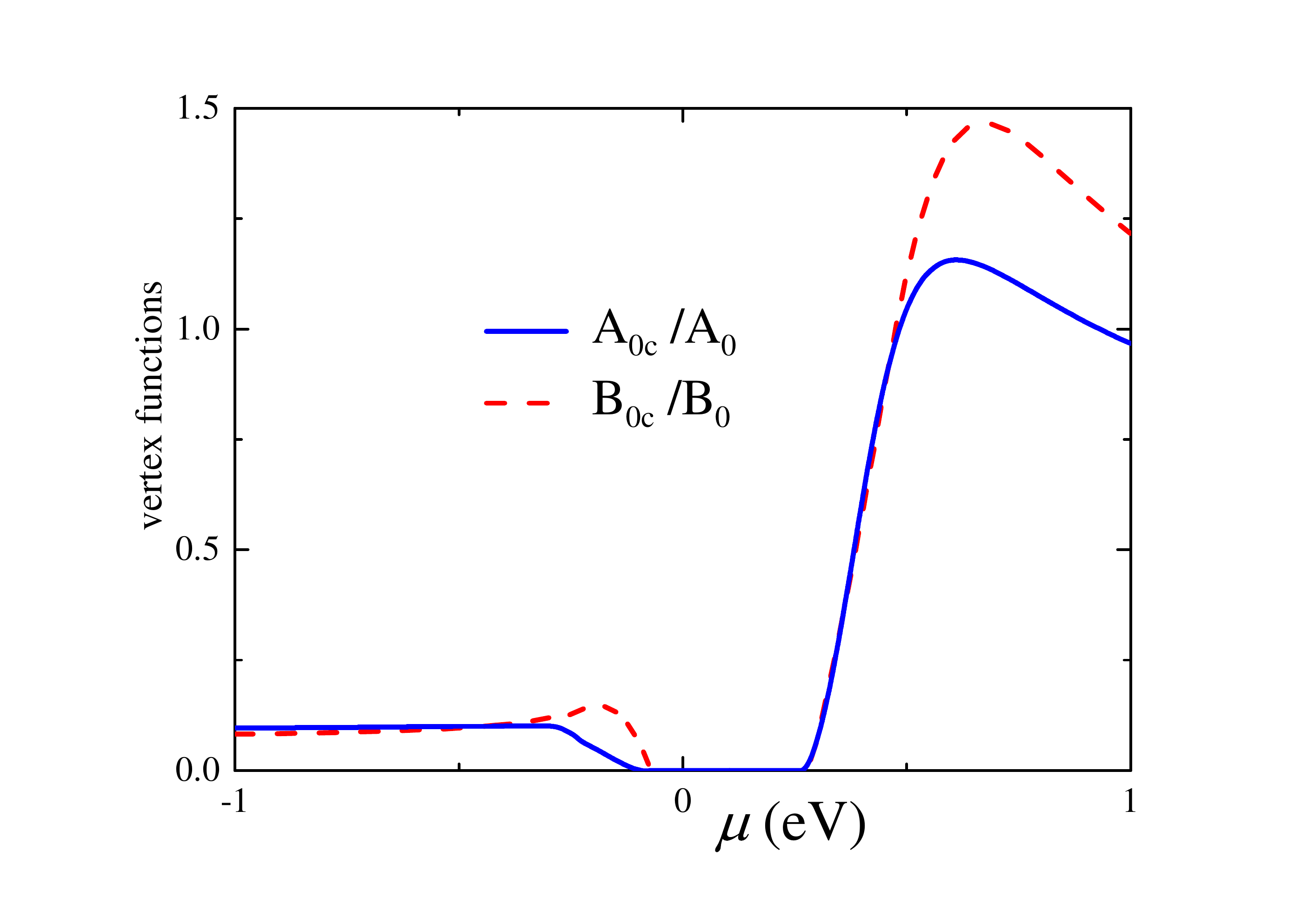}
    \caption{Vertex functions $A_{0c}$ (red line) and $B_{0c}$ (blue dashed line) as a functions of a chemical potential $\mu$.}
    \label{vertex}
\end{figure}

Vertex corrections are shown in Fig.~\ref{vertex}. As we can see, vertex corrected values $A_{0c}$ and $B_{0c}$ are comparable with the bare values $A_0$ and $B_0$ respectively. It is different from the case of the Rashba systems where vertex corrected values vanish~\cite{Inoue2004}. The values of the vertex corrections are almost independent of the chemical potential for the valence bands $\mu<0$. Vertex corrections are much larger for the conductivity bands $\mu>0$ than for the hole bands and have a maximum value near the gap. Vertex functions have a significant electron-hole asymmetry: vertex corrections are more significant for the conduction bands.  Origin of such asymmetry comes from the $C_1$ and $C_2$ terms that bring particle-hole asymmetry to the spectra of the Hamiltonian. Such terms reduce $k_F$ for the conductivity band $E_+(k_F)=\mu$ and increases $k_F$ for the valence bands $E_-(k_F)=\mu$. Here $E_{\pm}$ are given by Eq.~\ref{spectrum}. Vertex functions are proportional to the contribution from the linear part from the spectra which is larger for smaller $k_F$. Thus, vertex corrected values are larger for the conduction band. 

\section{Spin conductivity from the states at the Fermi surface}\label{fermi_surface}
Now we use the obtained results and Eq.~\eqref{spin_surface} to calculate the contribution to the spin conductivity from the states at the Fermi surface. On this way, we obtained that the term $\sigma_{{\alpha}\beta}^{II\gamma}$ vanishes exactly and we should compute the term $\sigma_{{\alpha}\beta}^{I\gamma}$ only.

Isotropic tensor components $\sigma_{xy}^{Iz}=-\sigma_{yx}^{Iz}$, $\sigma_{xz}^{Iy}=-\sigma_{yz}^{Ix}$ and $\sigma_{zx}^{Iy}=-\sigma_{zy}^{Ix}$ are the only terms that exists in the system. It can be presented as a sum of two terms
    \begin{eqnarray}\label{spinsum}
        \sigma_{{\alpha}\beta}^{I\gamma}=\sigma_{{\alpha}\beta}^{I\gamma bv} + \delta \sigma_{{\alpha}\beta}^{I\gamma}
    \end{eqnarray}
    where $\sigma_{{\alpha}\beta}^{I\gamma bv}$ is the contribution without vertex corrections, $\delta \sigma_{{\alpha}\beta}^{I\gamma}$ is the contribution that arises due to vertex corrections. Explicitly, that contributions can be written as
    \begin{eqnarray}\label{spin1}
    \sigma_{xy}^{Izbv}=-\sigma_0 \int k\,dk\,dk_z \frac {4 k^3A_0^2(C_2\Gamma_0-M_2\Gamma_z) }{E_{g+}E_{g-}},  \\ \nonumber
\delta    \sigma_{xy}^{Iz}=-\sigma_0 \int k\,dk\,dk_z \frac {4 k^3A_0A_{0c}(C_2\Gamma_0-M_2\Gamma_z) }{E_{g+}E_{g-}},  \\ 
    \end{eqnarray}
and
\begin{widetext}
    \begin{eqnarray}\label{spinz}
\sigma_{zx}^{Iy}=\sigma_{zx}^{Iybv} + \delta \sigma_{zx}^{Iy}, \\ \nonumber
\sigma_{zx}^{Iybv}=\sigma_{yz}^{Ixbv}=\sigma_0 \int k\,dk\,dk_z \frac {4 kA_0B_0 (\Gamma_0(M_1k_z^2-M_0)+\Gamma_z(C_0-\mu-C_1k_z^2))}{E_{g+}E_{g-}},  \\ \nonumber
\delta    \sigma_{zx}^{Iy}=-\sigma_0 \int k\,dk\,dk_z \frac {4 kA_{0c}B_0 (\Gamma_0(M_1k_z^2-M_0-M_2k^2)+\Gamma_z(C_0-\mu-C_1k_z^2+C_2k^2))}{E_{g+}E_{g-}},  \\ 
\delta    \sigma_{xz}^{Iy}=-\sigma_0 \int k\,dk\,dk_z \frac {4 kA_{0}B_{0c} (\Gamma_0(M_1k_z^2-M_0-M_2k^2)+\Gamma_z(C_0-\mu-C_1k_z^2+C_2k^2))}{E_{g+}E_{g-}},  \\ \nonumber
E_{g\pm}=(M_0+k_z^2M_1+k^2M_2\mp i\Gamma_z)^2-(C_0-\mu+C_1k^2+C_2k_z^2\pm i\Gamma_0)^2+A_0^2k^2+B_0^2k_z^2
\end{eqnarray}
\end{widetext}
where $\sigma_0= e/(2\pi)^3$. Note, that we use a different notation from the standard definition of the spin conductivity quanta $\sigma_0^z=e/(8\pi)$. 

\begin{figure}[t!]
    \centering
    \includegraphics[width=8.5cm, height=4.5cm]{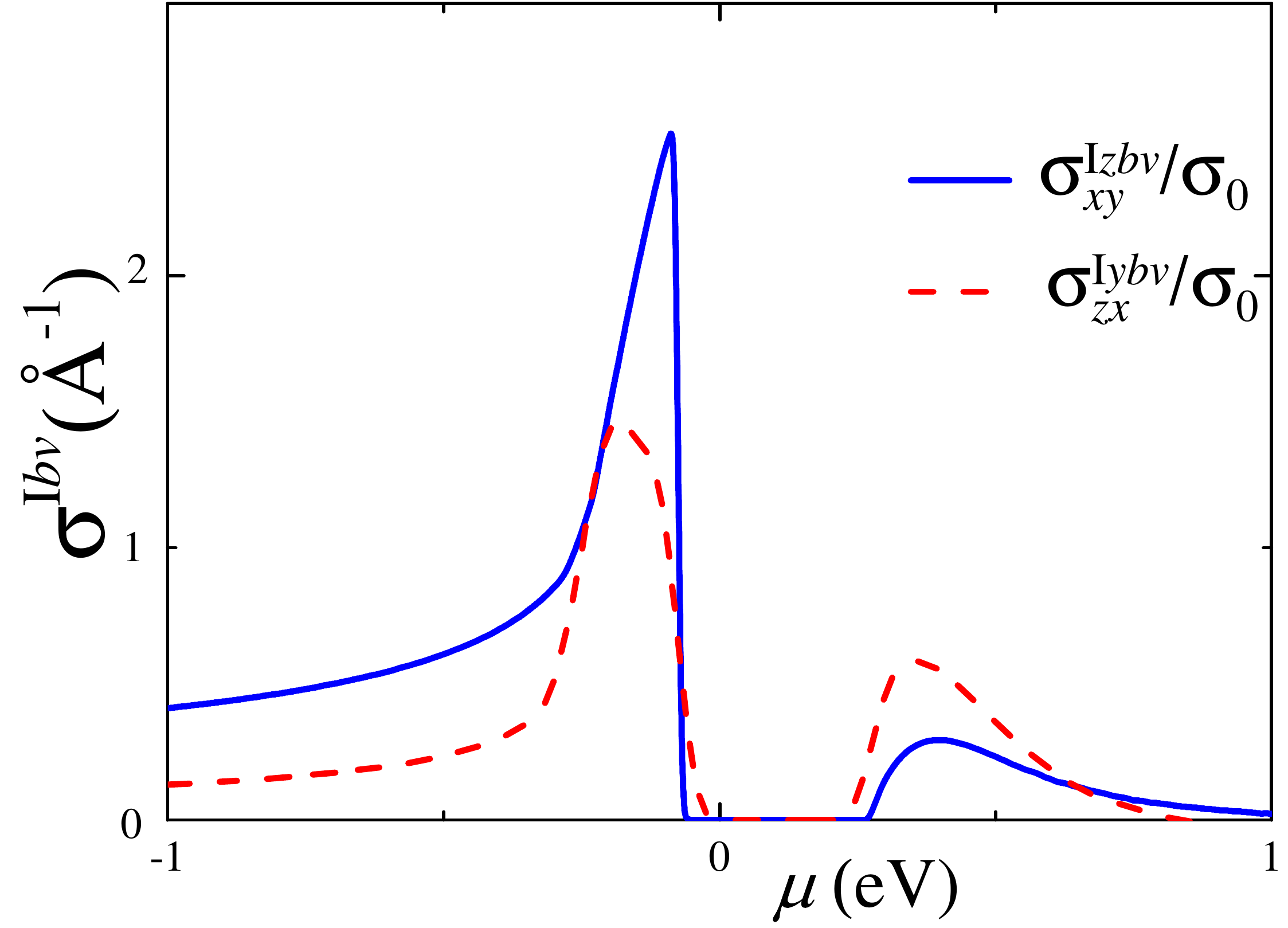}
    \caption{Bare spin conductivity contribution from the states at the Fermi surface $\sigma_{{\alpha}\beta}^{I\gamma bv}$ versus the chemical potential $\mu$. Blue line corresponds to $\sigma_{xy}^{Iz bv}$, red dashed line is $\sigma_{zx}^{Iy bv}=\sigma_{xz}^{Iy bv}$. }\label{surface1}
\end{figure}

\begin{figure}[t!]
    \centering
    \includegraphics [width=8.5cm, height=4.5cm]{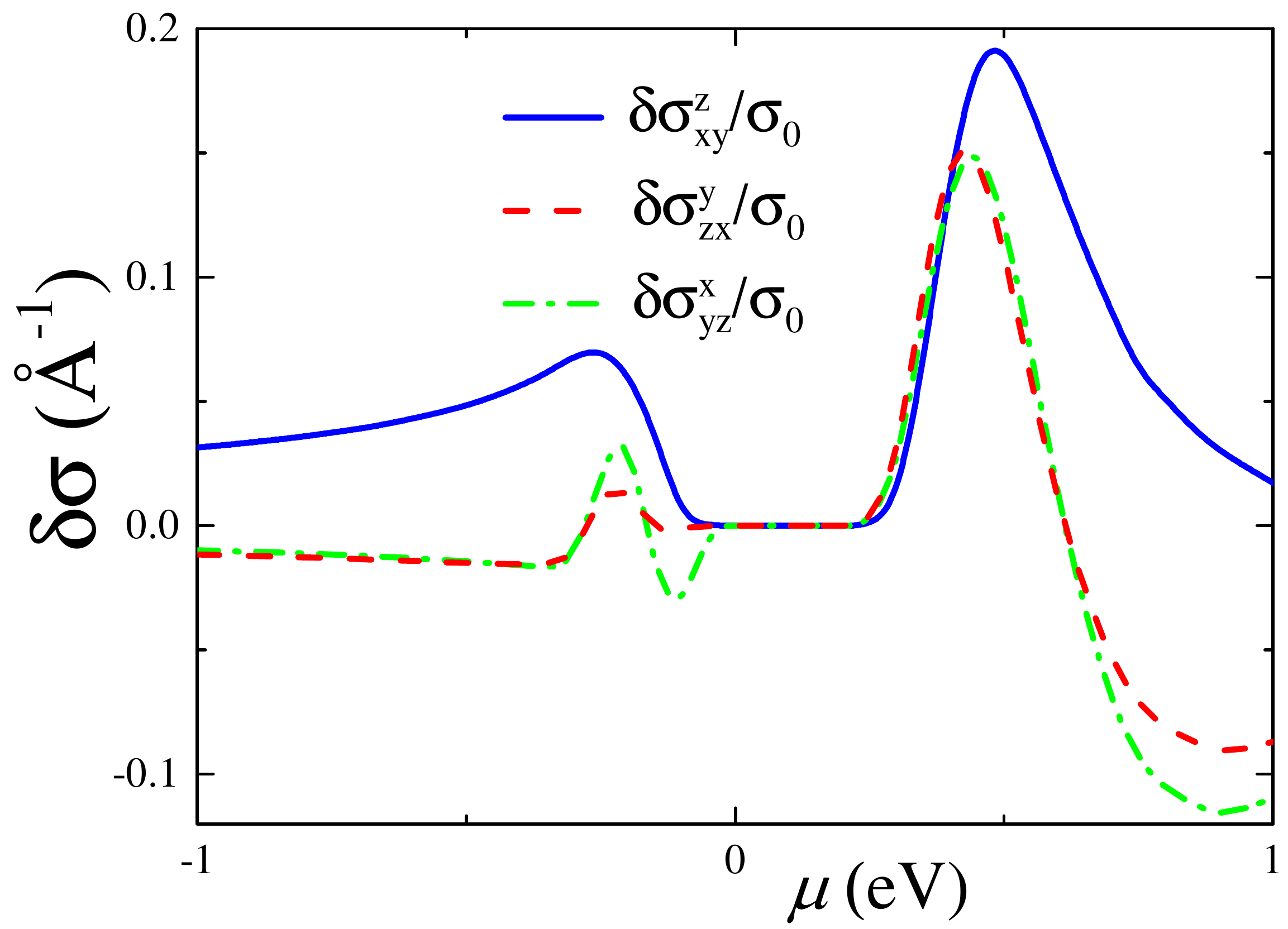}
    \caption{Spin conductivity contribution from the vertex corrections $\delta\sigma_{{\alpha}\beta}^{ I\gamma}$ versus the chemical potential $\mu$. Blue line corresponds to $\delta\sigma_{xy}^{Iz }$, red dashed line is $\delta\sigma_{zx}^{Iy}=\delta\sigma_{xz}^{Iy }$.}\label{surfacevc}
\end{figure}

We plot spin conductivity contributions as a function of the chemical potential at Figs.\ref{surface1} and \ref{surfacevc}. We can see those spin conductivity contributions take maximal values near the gap for the conductance bands. Also, spin conductivity contributions $\sigma_{xy}^{Iz}$, $\sigma_{zx}^{Iy}$ and $\sigma_{xz}^{Iy}$ have the similar values.  Spin conductivity's from vertex corrections $\delta\sigma_{zx}^{Iy bv}$ and $\delta\sigma_{xz}^{Iy bv}$ changes its sign due at some value of the chemical potential due to change of the sign of the orbital scattering $\Gamma_z$, see Fig.~\ref{scattering}.
\section{Spin conductivity from the filled states and total spin conductivity}\label{total}
We study Eq.~\ref{spin_filled} in a weak disorder limit. Isotropic tensor components $\sigma_{xy}^{IIIz}=-\sigma_{yx}^{IIIz}$, $\sigma_{xz}^{IIIy}=-\sigma_{yz}^{IIIx}$ and $\sigma_{zx}^{IIIy}=-\sigma_{zy}^{IIIx}$ are the only terms that exist in the system. After energy integration using constant damp approximation for the $\Gamma_0$ and $\Gamma_z$ we get that contribution to the spin conductivity from the filled states consists of two parts
\begin{equation} \label{spin2}
\sigma_{xy}^{IIIz}=\sigma_{xy0}^{z}-\sigma_{xy}^{Izbv},
\end{equation}
where $\sigma_{xy0}^{z}$ is the spin conductivity in a clean limit, $\sigma_{xy}^{Izbv}$ is given by Eq.~\ref{spin1}. Similar expression has been obtained for the anomalous charge Hall conductivity for the Dirac Hamiltonian~\cite{Sinitsyn2007,Nunner2007}.
Spin conductivity from the filled states in a clean limit can be expressed as~\cite{Sinova2004,Sinitsyn2004}
\begin{eqnarray}\label{sigmaIII}
\sigma_{{\alpha}\beta 0}^{\gamma}\!\!=\!\!e\!\!\!\! \sum\limits_{\mathbf{k},n\neq n'}\!\!\!(f_{n\mathbf{k}}\!-\!f_{n'\mathbf{k}})\!
\frac{ \textrm{Im}\langle u_{n'\mathbf{k}}|j_{\alpha}^{\gamma}|u_{n\mathbf{k}}\rangle\!\langle u_{n\mathbf{k}}|v_{\beta}|u_{n'\mathbf{k}}\rangle}{(E_{n\mathbf{k}}-E_{n'\mathbf{k}})^2}.\,\quad
\end{eqnarray}
Here $E_{n\mathbf{k}}$ is the energy of an electron in the $n$-th band with the momentum $\mathbf{k}$, $u_{n\mathbf{k}}$ is the corresponding Bloch vector, $\hat{H} u_{n\mathbf{k}} = E_{n\mathbf{k}} u_{n\mathbf{k}}$, $f_{n\mathbf{k}}$ is the Fermi distribution function corresponding to $E_{n\mathbf{k}}$ (which is the Heaviside step-function in the considered case of zero temperature), Im is for the imaginary part, $\langle ... \rangle$ is the scalar product here.
Using Eq.~\ref{sigmaIII} we obtain 
\begin{widetext}
\begin{eqnarray}\label{top_eq}
\sigma_{xy0}^{z}=\sigma_0 \int\limits_0^{\infty} (\theta(E_+)-\theta(E_-))k\,dkdk_z \frac {\pi k^2A_0^2 C_2}{2(A_0^2k^2+B_0^2k_z^2+(M_0+M_1k_z^2+M_2k^2)^2)^{3/2}}, 
\\
\sigma_{zx0}^{y}=\sigma_{xz0}^{y}=\sigma_0 \int\limits_0^{\infty} (\theta(E_+)-\theta(E_-))k\,dk\,dk_z \frac {\pi A_0B_0(M_0-M_1k_z^2)}{2(A_0^2k^2+B_0k_z^2+(M_0+M_1k_z^2+M_2k^2)^2)^{3/2}}, 
\end{eqnarray}
\end{widetext}
where $\theta(x)$ is Heaviside step function. Finite disorder has a little impact on this term in the spin conductivity: Heaviside step function $\theta(E_{\pm})$ is replaced by the normalized arctangent function $2/\pi\arctan E_{\pm}/\Gamma_0$. This substitution leads to the insignificant blurring of the spin conductivity near the gap for a small disorder.

\begin{figure}[t!]
    \centering
    \includegraphics [width=8.5cm, height=4.5cm]{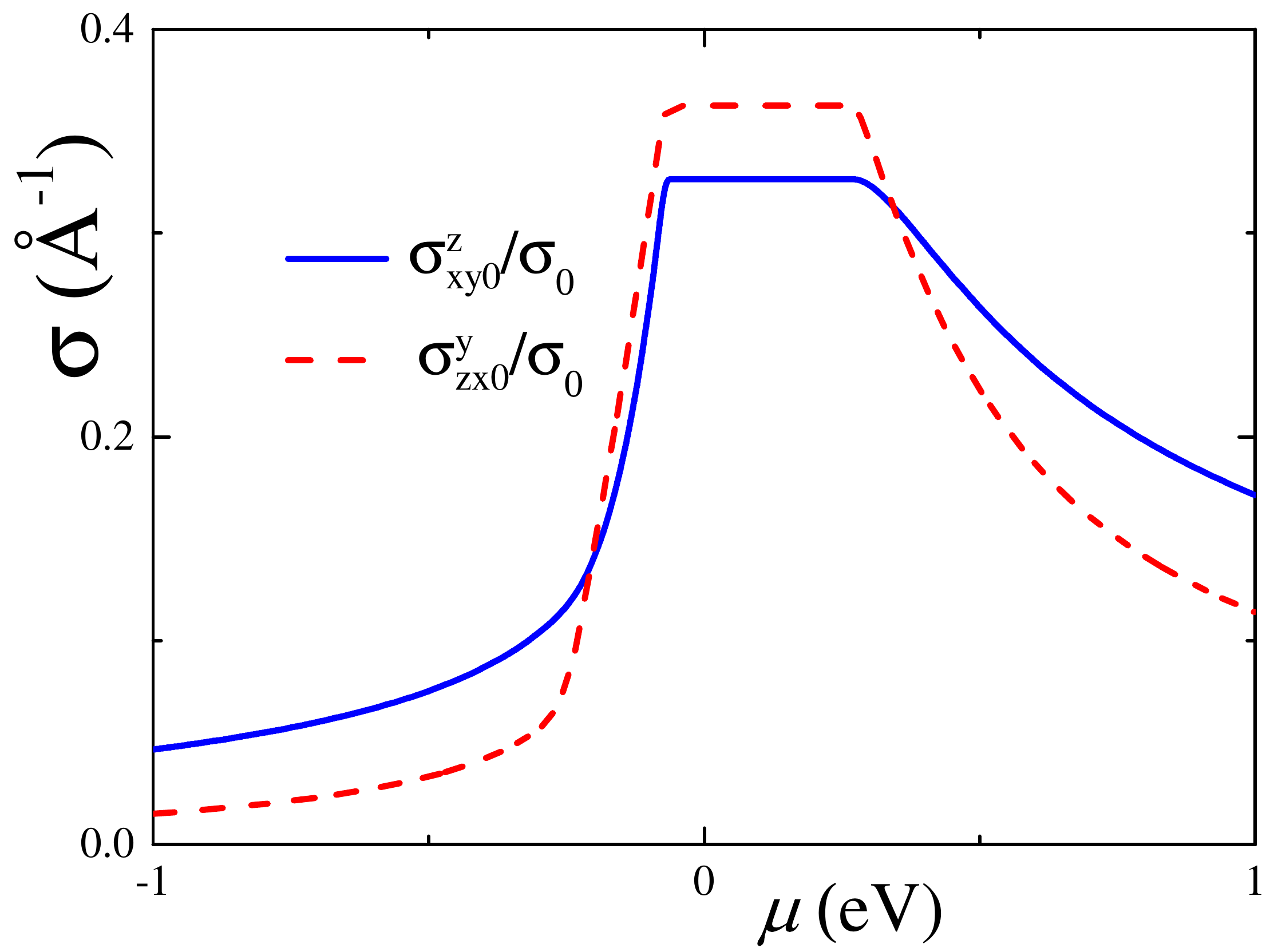}
    \caption{Spin conductivity contribution from the filled states in a clean limit $\sigma_{{\alpha}\beta0}^{\gamma}$ versus the chemical potential $\mu$. Blue line corresponds to $\sigma_{xy0}^{z}$, red dashed line is $\sigma_{zx0}^{y}=\sigma_{xz0}^{y}$.}\label{top1}
\end{figure}

Spin conductivity contribution from the filled states is shown in Fig.~\ref{top1}. We see that it is finite inside the gap. Outside the gap, this contribution to the spin conductivity decreases.

In case of small $B_0 \rightarrow 0$ we can get a simple formula for the spin conductivity for the gapped region
\begin{equation}\label{top_analytical}
\sigma_{xy}^{z}=\sigma_0 \frac{\pi^2 A_0 C_2}{4\sqrt{M_1 M_2^3}}.
\end{equation}
This formula reproduces accurate calculations of the Eq.~\ref{top_eq} quite well. 

We can see from Eqs.~\ref{spin1} and~\ref{spin2} that the term $\sigma_{\alpha\beta }^{I\gamma bv}$ cancels out from $\sigma_{\alpha\beta}^{I\gamma}$ and $\sigma_{\alpha\beta}^{III\gamma}$. So, only contributions to the spin conductivity from the vertex corrections and spin conductivity in a clean limit remain. 
We plot total spin conductivities $\sigma_{\alpha\beta}^{\gamma}=\delta \sigma_{\alpha\beta}^{I\gamma}+\sigma_{\alpha\beta0}^{\gamma}$ as a functions of chemical potential for the experimentally relevant parameters. Spin conductivity remains finite inside the gap and is the largest for the conduction band if chemical potential is not too large. Spin conductivity is suppressed for the valence bands in comparison with the spin conductivity for the conduction bands.  Spin conductivity changes its sign for the large values of the chemical potential. Out-of-plane spin conductivity $\sigma_{yz}^{x}$ is larger than in-plane spin conductivity $\sigma_{xy}^{z}$ inside the gapped region. However, in the metallic region, in-plane spin conductivity $\sigma_{xy}^{z}$ is larger and saturates slowly.     
\begin{figure}[t!]
    \centering
    \includegraphics [width=8.5cm, height=4.5cm]{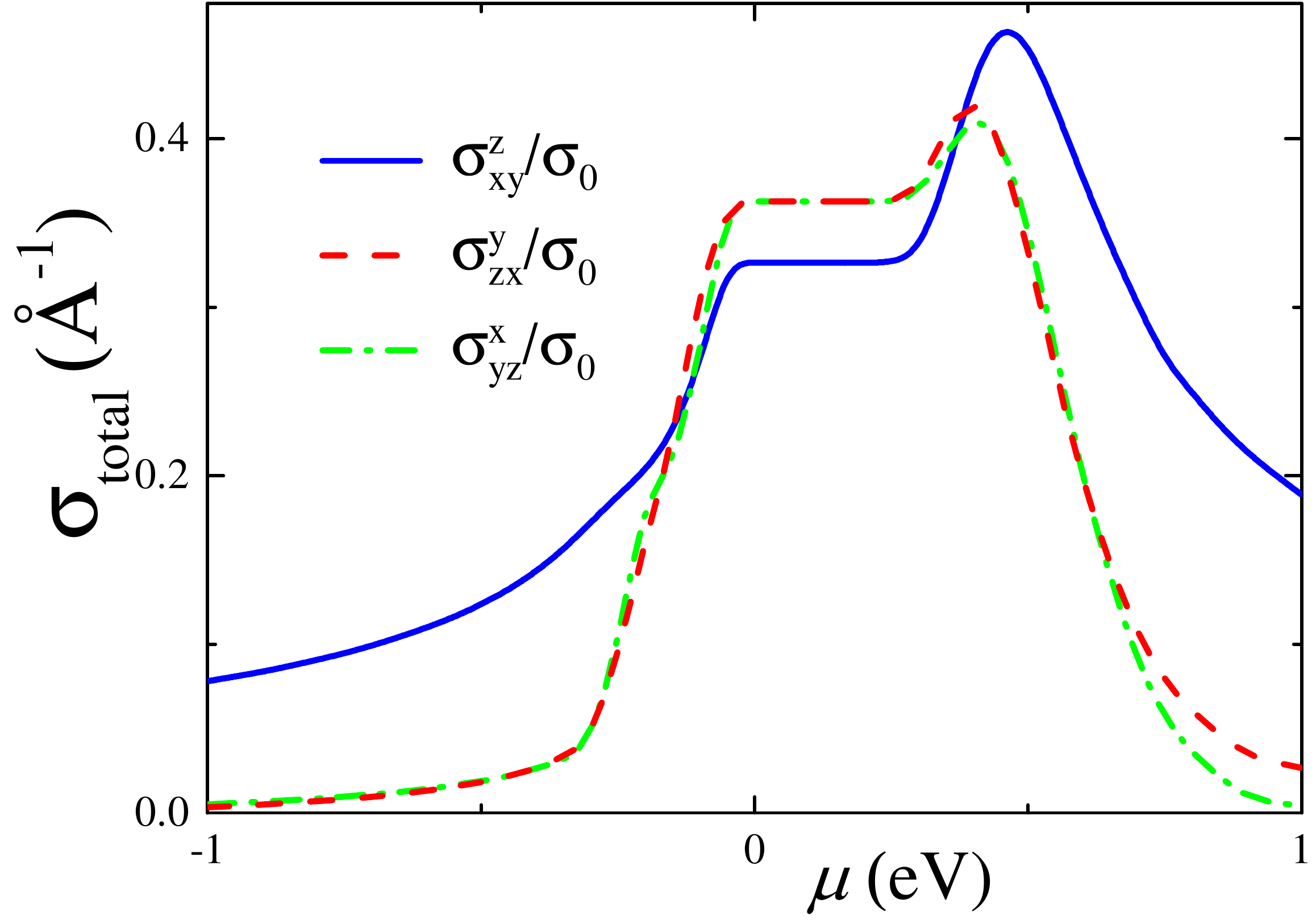}
    \caption{Total spin conductivities $\sigma_{xy}^{z}$ (blue line), $\sigma_{zx}^{y}$ (red dashed line) and $\sigma_{yz}^{x}$ (green dot dashed line) for Bi$_2$Se$_3$ versus the chemical potential $\mu$.}\label{exp}
\end{figure}
\begin{figure}[t!]
    \centering
    \includegraphics [width=8.5cm, height=4.5cm]{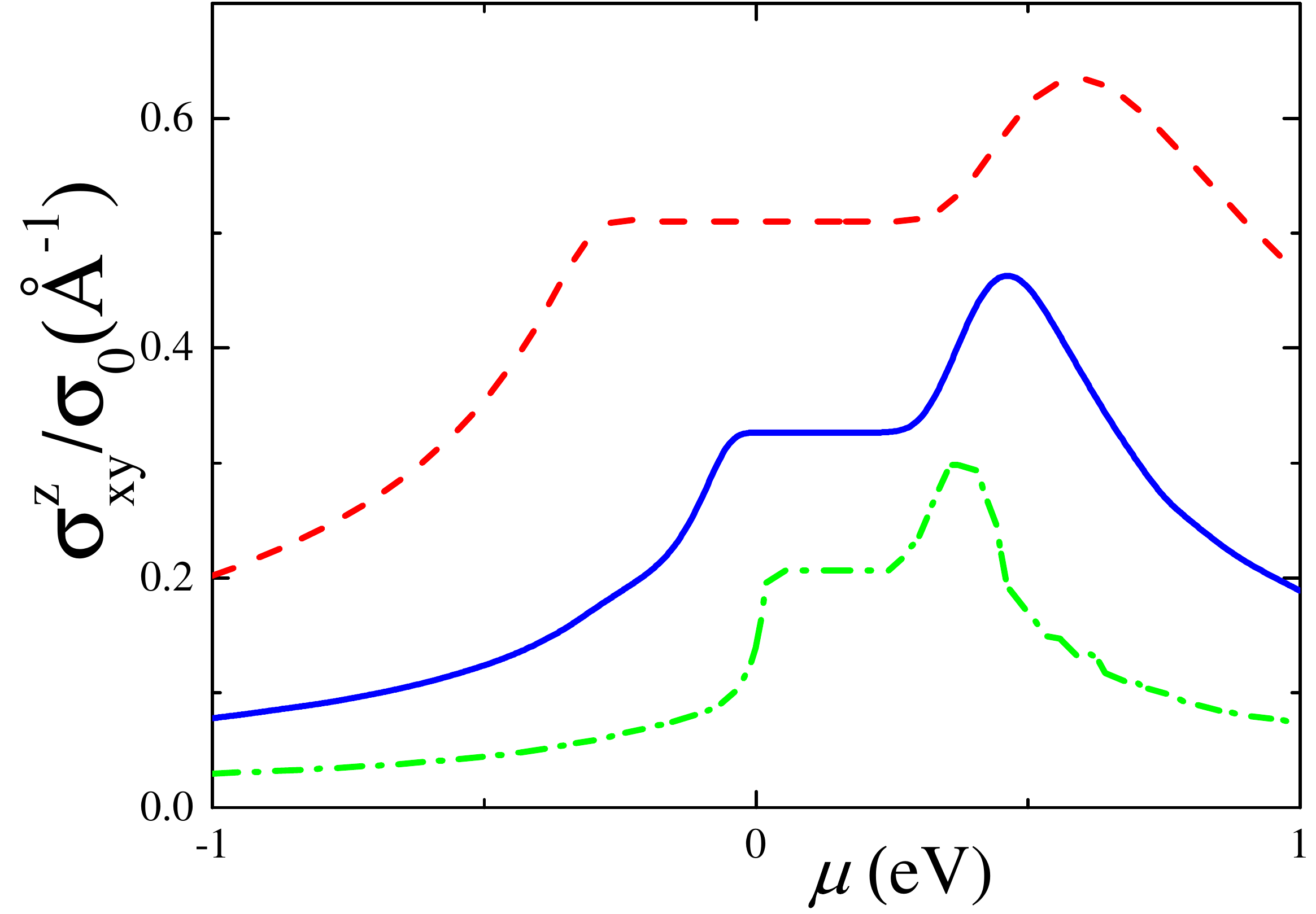}
    \caption{Total spin conductivity $\sigma_{xy}^{z}$ versus the chemical potential $\mu$ for the density functional theory parameters (blue line), for quadratic corrections ($C_1,C_2,M_1,M_2$) decreased by half (red dashed line) and for quadratic corrections ($C_1,C_2,M_1,M_2$) increased by half (green dot-dashed line).}\label{exp22}
\end{figure}

We plot spin conductivity $\sigma_{xy}^{z}$ for the different values of the quadratic in momentum corrections ($C_1,C_2,M_1,M_2$) to the Hamiltonian in Fig.~\ref{exp22}. Other components of the spin conductivity tensor have the same tendency concerning parameters change. These variations of the parameters can occur due to the chemical potential shift and structural distortions caused by the doping of the pristine topological insulator. Fermi velocity of the surface states of a topological insulator is closely tied with the value of the linear part of the bulk spectra~\cite{Liu2010}. Large deviations of the Fermi velocity of the surface states have been observed in the experiments: its value can be decreased by one\cite{Veldhorst2012} or even two orders\cite{Wolos2012} in comparison with the density functional theory parameters.

\section{Discussion}\label{discussion}
We get that maximal values of the spin conductivity of Bi$_2$Se$_3$ is approximately $\sigma_{xy}^{z} \sim \hbar/2e \, 10^4 \Omega^{-1}m^{-1}$ and $\sigma_{xz}^{y} \sim \hbar/2e \, 0.8 \cdot10^4 \Omega^{-1}m^{-1}$. These values are approximately one order of the amplitude smaller than it was measured in the experiment\cite{Mellnik2014} $\sigma_{||}=\hbar/2e \, 1.2 \cdot10^5 \Omega^{-1}m^{-1}$ and $\sigma_{\perp}=\hbar/2e \, 10^5 \Omega^{-1}m^{-1}$. As we can see from Eq.~\ref{top_analytical} the value of the spin conductivity is determined not only by the leading linear term but also by the next order corrections to the Hamiltonian. The values of such small terms are correct only near the vicinity of the $\Gamma$ point~\cite{Liu2010}. We argue that such a discrepancy between experimental data and our predictions occurs due to different values of the parameters in the real materials in comparison with the parameters calculated from the density functional theory. For the real materials, different values of the parameters can be realized. Upon doping Fermi surface transforms from the spherical to the cylindrical one~\cite{Lahoud2013}. This evolution occurs due to the change of the quadratic corrections to the spectra~\cite{Hao2017}. If we consider that all quadratic in momentum terms $C_1,C_2,M_1,M_2$ are one order of amplitude smaller we get that spin conductivity is almost one order of the amplitude larger, $\sigma_{xy}^{z} \sim \hbar/2e \, 0.8 \cdot 10^5 \Omega^{-1}m^{-1}$ and $\sigma_{xz}^{y} \sim \hbar/2e \, 0.7 \cdot10^4 \Omega^{-1}m^{-1}$, that is in agreement the experimental results.

Another explanation of the large spin conductivity is that surface states of topological insulators carry a large spin current. However, as we have calculated previously, surface states of the one side of the film contribute to the spin conductivity around~\cite{Akzyanov2019,Akzyanov2019a} $\sigma_{xy}^{z}\sim 2 \cdot 10^{-5} \Omega^{-1}$. This value is comparable to the contribution to the spin conductivity from the bulk states of one layer $\sigma_{xy}^{z}\sim 3 \cdot 10^{-5} \Omega^{-1}$.  Thus, a major contribution to the spin conductivity comes from the bulk states for thick samples.  If there are more than five layers, we can be sure that almost all spin current comes from the bulk states. 

Applying voltage along $x$ direction generates a spin current in $y$ direction with $z$ spin as well as spin current in $z$ direction with the $y$ spin. Thus, spin current in topological insulators will inevitably affect magnetization in in-plane and out-of-plane directions. There is no Rashba-Edelstein effect for the bulk states due to inversion symmetry in the bulk of topological insulators. Thus, damping and anti-damping torques for the bulk states are generated by the spin Hall effect solely. However, for the surface states inversion symmetry is broken and non-equilibrium spin density can be generated~\cite{Ghosh2018}. Also, doping of the sample can lead to the formation of two-dimensional Rashba gas near the surface of topological insulator~\cite{King2011}. Such two-dimensional gas does not contribute to the spin Hall conductivity due to vertex cancellation~\cite{Inoue2004,Raimondi2005}, but contribute to the Rashba-Edelstein effect. These contributions do not scale with the thickness of the sample and can be neglected for thick films. Thus, only spin Hall currents of the bulk states have a significant effect on the non-equilibrium spin properties of thick samples of topological insulators.


In our previous work~\cite{Akzyanov2019}, we missed the second term from Eq.~\ref{spin2} for the contribution to the spin conductivity from the filled states. This term leads to the cancellation of the contribution to the spin conductivity from the bulk states for a Rashba model. Together with the vertex cancellation spin conductivity vanishes for the Rashba model~\cite{Dimitrova2005}. Thus, the Rashba model cannot describe the significant spin conductivity of the bulk states of the topological insulator. Our omission does not qualitatively change previous results for the anisotropic terms of the spin conductivity and the spin conductivity of the surface states since vertex cancellation is absent for these cases~\cite{Akzyanov2019}.

We get that even for the insulating state finite non-quantized value of the spin conductivity exists. Spin conductivity in the insulating phase is determined by Eq.~\ref{top_eq}. If spin is conserved then this formula corresponds to the integral of the spin Berry curvature over the compact manifold that is quantized\cite{Dayi2015}. If spin is not conserved then spin conductivity is not quantized. We get a similar result previously for surface states in a thin film of topological insulator~\cite{Akzyanov2019a}. This phase is especially interesting for practical applications since topological insulators with a large bulk gap with very low charge concentrations are available~\cite{Kim2012}. Experimentally, spin conductivity can be measured by the spin-transfer torque effect~\cite{Ralph2008,Sinova2015}. In the insulating phase, not only surface states contribute to the charge current, but the whole bulk contributes to the spin current. Thus, the conversion of the charge current into the spin current would be extremely efficient for a thick sample. 

Band inversion in topological insulators leads to the non-trivial topology of the band structure. For our model, it leads to the finite spin conductivity inside the gap and the absence of the vertex cancellation of the spin conductivity. So, we can expect similar results for other topological materials with the inversion of the bands.

We neglected hexagonal warping since its value is small for the Bi$_2$Se$_3$~\cite{Kuroda2010}. However, for Bi$_2$Te$_3$ and other compounds, hexagonal warping can have a significant effect on the electron and spin properties\cite{Fu2009}. Similar to the surface states, hexagonal warping generates additional components in the spin conductivity tensor for the bulk states as well\cite{Akzyanov2019}. 

We found that the value of the disorder has a weak influence on the value of the spin conductivity in case of the weak scattering limit. Such a result has been obtained before for the Rashba model~\cite{Inoue2004,Akzyanov2019}. We can speculate that this is a common feature for the spin conductivity due to the absence of the Drude peak. At large enough disorder ladder approximation for the calculation of the self-energy and vertex corrections fails and more advanced techniques are required.

 Here we want to discuss the influence of the Hamiltonian parameters on the value of the spin conductivity.
Value of the spin conductivity term $\sigma_{xy}^z$ is mainly determined by the interplay of the linear part of the in-plane spectra $A_0$, in-plane quadratic term $M_21/2m_e^*$ that corresponds to the inverse effective carrier mass, in-plane quadratic term $C_2$ term that brings particle-hole asymmetry to the effective carrier mass and out-of-plane quadratic term $M_1$. Increase of the particle-hole asymmetry $C_2$ increases spin conductivity as well as increase of the effective in-plane carrier mass $\propto 1/M_2$. Increase of $M_1$ decreases $\sigma_{xy}^z$. Linear part $B_0$ and quadratic part $C_1$ that corresponds spectra along $k_z$ direction has a little influence on $\sigma_{xy}^z$. Values of the spin conductivity components $\sigma_{zx}^y$ and $\sigma_{yz}^x$ are mainly determined by $A_0$, $B_0$, $M_1$ and $M_2$. Increase of $A_0$, $B_0$ and $M_1$ increases $\sigma_{zx}^y$ and $\sigma_{yz}^x$ terms of the spin conductivity while increase of the in-plane quadratic terms $M_1$ decreases them.

 Our calculations are performed at zero temperature but can be easily generalized for a finite temperature. In case of contributions to the spin conductivity at the Fermi surface instead of $\sigma_{xy}^z(\mu)$ we should use $\delta \sigma_{xy}^z(T)=\int \frac{\partial f(E-\mu)}{\partial E}\, \delta \sigma_{xy}^z(E) dE$, where $f(E)$ is the Fermi distribution. In case of the contribution from the filled states, given by Eq.~\ref{top_eq}, finite temperature substitutes Heaviside step functions by Fermi distribution that leads to the small blurring of the spin conductivity. 

To sum up, transverse spin Hall currents in perpendicular directions with comparable amplitudes arise when the voltage is applied. Finite spin conductivity with the non-universal value exists in the gapped region. Intrinsic contribution to the spin conductivity from the filled states is comparable to the contribution from the vertex corrections. Our results are in an agreement with the experimental data. 

\section*{Acknowledgements}

RSA acknowledges support by Russian Science Foundation (Grant N. 17-12-01544) and Foundation for the Advancement of Theoretical Physics and Mathematics “BASIS”.

    \bibliographystyle{apsrev4-1}
    \bibliography{bib_hw}

\end{document}